\RequirePackage{fix-cm}

\documentclass[a4paper,11pt]{article}
\usepackage{jheppub}
\usepackage[T1]{fontenc} 

\newcommand{\ee}{e^+e^-}
\newcommand{\emu}{e^+\mu^-+e^-\mu^+}

\newcommand{\mumu}{\mu^+\mu^-}

\title{On the excess in the inclusive $W^+W^-\rightarrow
  l^+l^-\nu\bar\nu$ cross section}

\author[a]{Pier~Francesco~Monni}
\author[a,b]{Giulia Zanderighi}

\affiliation[a]{Rudolf Peierls Centre for Theoretical Physics,University of
  Oxford,1 Keble Road, \\Oxford OX1 3RH, UK}
\affiliation[b]{CERN,
  Theory Division, CH-1211 Geneva 23, Switzerland}

\emailAdd{pier.monni@physics.ox.ac.uk}
\emailAdd{giulia.zanderighi@cern.ch}

\abstract{
  In this note we analyse the excess in the $W^+W^-$ {\it inclusive}
  cross section recently measured at the LHC. We point out that in
  fact for the ATLAS {\it fiducial} cross sections there is no excess
  in the measurements compared to the NLO QCD predictions. We also
  argue that higher order effects to the fiducial cross section are
  small, and tend to cancel each other, hence the inclusion of NNLO
  and NNLL corrections will not modify this agreement
  significantly. We find that at 8 TeV a substantial part of the
  disagreement with the NLO prediction for the total cross section
  observed by ATLAS is due to the extrapolation carried out with {\tt
    POWHEG}.}

\begin{document}
\maketitle
\flushbottom

\section{Introduction}
The inclusive $W^+W^-$ cross section at $7$ and $8\,{\rm TeV}$ has
been measured recently by both the
ATLAS~\cite{ATLAS:2012mec,ATLAS:2013confnote} and CMS
collaborations~\cite{Chatrchyan:2013yaa,Chatrchyan:2013oev}.  All
measurements show a systematic tension when compared to
next-to-leading order (NLO) QCD calculations.
The disagreement was first observed in the 7 TeV data, and then it
increased at 8 TeV, where the tension with the NLO predictions reaches
the $2-2.5\,\sigma$ level.
The small uncertainties quoted for the NLO calculations suggest that
higher-order QCD corrections can not change this pattern.
This triggered a lot of interest and a number of models were suggested
to explain this excess in terms of new light states, see
e.g. refs.\cite{Feigl:2012df,Curtin:2012nn,Jaiswal:2013xra,Rolbiecki:2013fia,Curtin:2014zua}.
However, before discussing any hint for New Physics, one needs to fully control the uncertainty associated with the Standard Model prediction. 

One issue that arises, is that
refs.~\cite{ATLAS:2012mec,ATLAS:2013confnote,Chatrchyan:2013yaa,Chatrchyan:2013oev}
quote the inclusive cross section, obtained by extrapolating the
measured fiducial cross section through data driven Monte Carlo (MC)
acceptances. One reason for quoting the inclusive cross-section is
that it is independent of the experimental setup, hence it is possible
to make statements about whether two measurements by ATLAS and CMS are
in agreement, while fiducial cross-sections are different for the two
experiments.
Nevertheless, the extrapolation from the fiducial to the inclusive
phase space relies on a Monte Carlo simulation, and thus obviously it
depends on the generator used.
For instance, a generator that systematically underestimates the
fiducial cross-section would lead to an overestimation of the
resulting "measured" inclusive cross section.  Hence, a comparison to
theory should be made first in the fiducial region, for which
experimental data is available, before extrapolating the result to the
fully inclusive phase space.  A possible way to compare the
measurements of both experiments avoiding a big extrapolation to the
total inclusive phase space would be to extrapolate the ATLAS fiducial
measurement from the CMS one, and vice versa. This extrapolation would
still depend on the Monte Carlo generator used, but it would involve
two cross sections that are typically of the same order of magnitude.

In this note we compare the measured fiducial cross sections to the
NLO predictions and find that there is no sizable tension between the
two (at the 1-$\sigma$ level).
We then study the effect of the extrapolation from the fiducial region
to the inclusive one.  We find that the Monte Carlo acceptance
computed with {\tt POWHEG} overestimates the reduction due to the
fiducial phase space cuts, leading to a larger total cross section
when the extrapolation from the fiducial to the inclusive phase space
is carried out.  We study the source of the reduction in the Monte
Carlo prediction, and discuss the possible impact of higher-order
corrections.
At the moment, only ATLAS published the measured fiducial cross
sections at 7 and 8 TeV. Since the larger disagreement is observed in
the 8 TeV data, we focus on the latter measurement for the three
leptonic channels, $\ee$, $\mumu$ and $\emu$.

\section{Comparison to Next to Leading Order prediction}

We present here theory predictions for the fiducial cross section as
defined by the ATLAS experiment at a centre-of-mass energy of 8
TeV~\cite{ATLAS:2013confnote}. The relative fiducial cuts are
summarized in Table~\ref{tab:fiducials}.
Our analysis equally applies to the 7 TeV case, for which we find similar conclusions. 

\begin{table}[htp]
\centering
\footnotesize
\begin{tabular}{|c|}
\hline
8 TeV fiducial region\\
\hline\hline\\
$p_t>25 (20)$ GeV for the
leading (subleading) lepton and charged leptons separated by $\Delta R>0.1$\\\\
muon pseudorapidity $|y|<2.4$ {\rm and}  
electron
pseudorapidity $|y|<1.37$ or $1.52<|y|<2.47$\\\\
no jets (anti-$k_t$~\cite{Cacciari:2008gp}, $R=0.4$) with $p_t>25$
GeV and $|y|<4.5$, separated from the electron
by $\Delta R>0.3$\\\\
$m_{ll'}>15,15,10$ GeV  and  $|m_{ll'}-m_Z|>15,15,0$ GeV for $ee$, $\mu\mu$, and $e\mu$, respectively\\\\
$p_{t,{\rm Rel}}^{\nu+\bar\nu}>45,45,15$ GeV and $p_{t}^{\nu+\bar\nu}>45,45,20$ GeV  for $ee$, $\mu\mu$, and 
$e\mu$, respectively\\
\\\hline
\end{tabular}
\caption{Fiducial volume, as defined by the ATLAS collaboration, at 8 TeV. For a detailed definition of all variables see~\cite{ATLAS:2013confnote}.\label{tab:fiducials}}
\end{table}

It is instructive to first compare the fiducial measurements to the
next-to-leading order results from {\tt MCFM
  6.6}~\cite{Campbell:1999ah}, including the formally
next-to-next-to-leading order (NNLO) contribution due to $gg\to
W^+W^-$~\cite{Campbell:2011cu}. We work in the $G_\mu$ scheme with the
electroweak parameters $M_Z=91.1876$\,GeV, $\Gamma_Z=2.4952$\,GeV,
$M_W=80.398$\,GeV, $\Gamma_W=2.1054$\,GeV, and $G_F=1.16639\times
10^{-5}$\,GeV$^{-2}$.  The branching ratio of $W$ into leptons is
taken from the Particle Data Book~\cite{Beringer:1900zz}, namely ${\rm
  Br}(W\rightarrow l \nu)=0.108$.\footnote{We stress that the
  experimental errors in the electroweak parameters are an important
  source of theory uncertainty. The sole variation of the $W$-boson
  width $\Gamma_{\rm W}$ within its error, as quoted by the PDG
  $\Gamma_{\rm W} = 2.085\pm 0.042$ leads to a $\sim$ 8.5\% variation
  in the total cross section, since it scales as $\sigma_{WW}\sim
  1/(\Gamma_{\rm W} M_{\rm W})^2$. However, the Standard Model
  calculation of the $W$ width has a much smaller uncertainty.  } We
use the MSTW2008nlo parton densities~\cite{Martin:2009iq} and central
renormalization and factorization scales $\mu_R=\mu_F=M_{\rm 4l}$,
where $M_{\rm 4l}$ is the invariant mass of the leptonic system. The
scale uncertainty is obtained by varying both scales by a factor of
two in either direction and keeping $\mu_R=\mu_F$.  With this setup we
get a total inclusive $W^+W^-$ cross section of
$53.6^{+1.4}_{-1.0}$~pb ($56.5^{+1.5}_{-1.1}$~pb) excluding
(including) Higgs mediation.  For consistency, the quoted Higgs
mediated cross section has also been computed with MCFM.  On the other
hand, ATLAS finds $71.4^{+1.2}_{-1.2}({\rm stat.})^{+5.0}_{-4.4}({\rm
  syst.})^{+2.2}_{-2.1}({\rm lumi.})$~pb.

The measured fiducial cross sections at 8 TeV are reported in the
first column of Table~\ref{tab:fidATLAS8vsNLO-MSWT-m(WW)}, together
with the total NLO predictions in the fourth column. The latter is
given by the sum of the cross-section without Higgs mediation and the
Higgs mediated contribution, as reported in the second and third
column, respectively.
The quoted theory uncertainty does not account either for the PDF
error, that was found to be at $5\%$ level~\cite{ATLAS:2013confnote},
or for any interference effect.

\begin{table}[htp]
\centering
\footnotesize
\begin{tabular}{|c||c||c|c|c|}
\hline
     &  ATLAS @ 8 TeV &$pp \to l^+l^-\nu\bar\nu$ &   $pp\to H\to  l^+l^-\nu\bar\nu$ & total \\
 \hline
$\emu$ &  $377.8^{+6.9}_{-6.8}{\rm (stat.)}^{+25.1}_{-22.2}{\rm (syst.)}^{+11.4}_{-10.7}{\rm (lumi.)} $  &$332.4^{+4.7}_{-2.3}$&$9.8^{+0.0}_{-1.2}$&$342.2^{+4.7}_{-2.6}$\\
$\ee$ & $68.5^{+4.2}_{-4.1}{\rm (stat.)}^{+7.7}_{-6.6}{\rm
  (syst.)}^{+2.1}_{-2.0}{\rm (lumi.)} $ &$63.7^{+0.8}_{-0.4}$&$2.2^{+0.0}_{-0.2}$&$65.9^{+0.8}_{-0.4}$\\
$\mumu$ &$74.4^{+3.3}_{-3.2}{\rm (stat.)}^{+7.0}_{-6.0}{\rm
  (syst.)}^{+2.3}_{-2.1}{\rm (lumi.)} $  &$69.3^{+0.9}_{-0.4}$&$2.4^{+0.0}_{-0.2}$&$71.7^{+0.9}_{-0.5}$\\
\hline 
\end{tabular}
\caption{ATLAS fiducial cross-sections in fb at 8 TeV 
  (1st column) and two processes that contribute to $W^+W^-$
  production (2nd and 3rd column). The last column contains the sum of the previous two contributions. The theory predictions are obtained with {\tt MCFM} with central renormalisation and factorisation scales set 
  to the leptonic system invariant mass, and using the MSTW2008nlo PDF set. 
  The (formally NNLO) $gg\rightarrow
  W^+W^-$ channel is included. Following ref.~\cite{ATLAS:2013confnote}, all quoted numbers do not include  electrons or muons coming from $\tau$ decays. 
  \label{tab:fidATLAS8vsNLO-MSWT-m(WW)}}
\end{table}

The quoted theoretical uncertainties obtained by varying
renormalisation and factorisation scales are tiny, at the level of
just about 1-1.5\%.
However, the definition of the fiducial volume involves a veto on jets
with a large transverse momentum (see Table~\ref{tab:fiducials}).
It is well-known that, in the presence of a jet-veto, fixed-order
calculations typically underestimate the true theoretical uncertainty,
and a more sophisticated procedure should be used in order to assess
the uncertainty~\cite{Stewart:2011cf,Banfi:2012jm}. Since at the
moment the by far dominant uncertainty is the systematic one, and
since several theoretical improvements are now possible on the theory
numbers quoted above (inclusion of NNLL resummation
effects~\cite{Jaiswal:2014yba,Meade:2014fca} matched to exact NNLO
predictions~\cite{Gehrmann:2014fva}, and NNLL threshold resummation in
the total cross section~\cite{Dawson:2013lya}), we do not try to
quantify the theory error in a more precise way. Yet, we observe that
there is {no sizable tension} between NLO theory and experiment within
the large systematic uncertainties. The level of agreement depends on
the leptonic channel considered, and it is always at the one $\sigma$
level, or better.

In the following section we rather study how much the Monte Carlo
prediction differs from the pure NLO one for the fiducial cross
section.  The ATLAS analysis at 8 TeV use the {\tt POWHEG}, so we
consider this generator in the following section.

\section{Resummation and Parton Shower effects}

In the present section we investigate the effect of parton shower and
hadronisation on the fiducial cross section using the {\tt POWHEG
  BOX}~\cite{Nason:2004rx,Frixione:2007vw,Alioli:2010xd} (using its
default settings), where the NLO prediction for $W^+W^-$ production is
matched to a parton shower~\cite{Melia:2011tj,Nason:2013ydw}.  Events
are showered with {\tt Pythia }6.4.28~\cite{Sjostrand:2006za}, Perugia
tune 350~\cite{Skands:2009zm} (unless otherwise stated), and
hadronisation and underlying event effects are included.  To estimate
the impact of parton shower and hadronisation we compare the latter
prediction to the NLO result obtained within the {\tt POWHEG}
programme itself. Unlike in the previous section, and for the sake of
simplicity, we do not include the $gg$-initiated channel here since it
is not implemented in the {\tt POWHEG
  BOX}. 

From Table~\ref{tab:fidATLAS8vsNLO-MSWT-m(WW)-PWG} it is evident that
the shower and hadronisation reduce the fiducial cross section
systematically in all channels and at both energies by about
9-11\%. The uncertainties with {\tt POWHEG} are obtained by varying
$\mu_R$ and $\mu_F$ by a factor of two in either direction around
$M_{4l}$, while keeping $\mu_R=\mu_F$.
\begin{table}[htp]
\centering
\footnotesize
\begin{tabular}{|c||c||c|c|}
\hline
      $pp\to  l^+l^-\nu\bar\nu$ (no gg) & ${\rm {\tt POWHEG} \,(hadron)}$ & ${\rm {\tt POWHEG} \,(NLO)}$&ratio\\
 \hline
$\emu$ &$295.2^{+4.8}_{-2.8}$&$323.0^{+6.0}_{-6.5}$&$0.91$\\
$\ee$ &$54.8^{+1.7}_{-0.7}$&$61.5^{+1.2}_{-1.3}$&$0.89$\\
$\mumu$  &$59.5^{+1.7}_{-0.2}$&$66.9^{+1.3}_{-1.6}$&$0.89$\\
\hline 
\end{tabular}
\caption{Comparison between {\tt POWHEG}  at hadron level and its NLO prediction with fiducial cuts of ATLAS at 8 TeV. Cross-sections are given in fb. Neither the $gg$-initiated process nor the Higgs-mediated one are included.
 The central scales are set  to the leptonic system invariant mass and the MSTW2008nlo PDF
 are used. Uncertainties are obtained as explained in the text.
 \label{tab:fidATLAS8vsNLO-MSWT-m(WW)-PWG}}
\end{table}
In the presence of a parton shower, competing effects will change the
hardest jet transverse momentum, e.g. events will typically have more
jets, while MPI and out of jet radiation can affect the hardest jet
transverse momentum. So it is reasonable to expect differences between
the pure NLO and {\tt POWHEG} prediction, especially given that the
hardest jet kinematics is described at leading order only in both
cases.
Yet, the observed 9-11\% effect is larger than what one would expect
for a process that at Born level is quark-initiated and involves no
final state QCD hard jets.

\begin{figure}[htp]
  \centering
  \includegraphics[width=0.5\columnwidth]{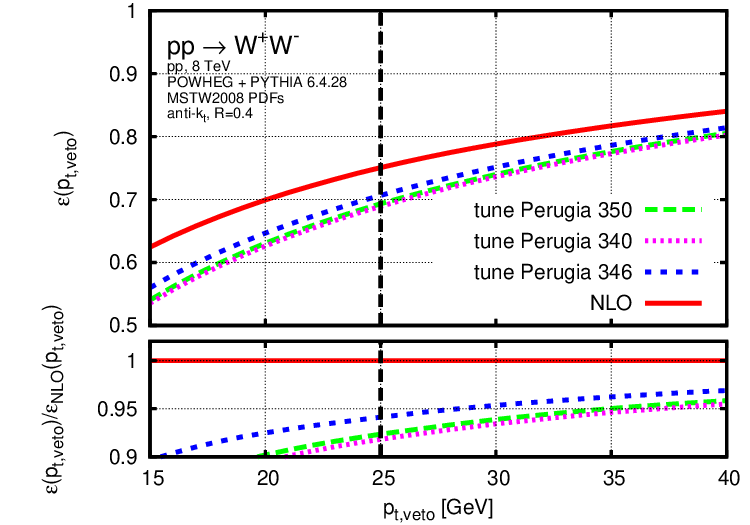}
  \caption{Jet-veto efficiency for $W^+W^-$ production obtained with
    the {\tt POWHEG BOX} both at pure NLO and matched to parton shower
    simulated with {\tt Pythia} 6.4.28. For comparison, different
    tunes are shown. }
  \label{fig:eff-PWG}
\end{figure}

\begin{figure}[htp]
  \centering
  \includegraphics[width=0.33\columnwidth]{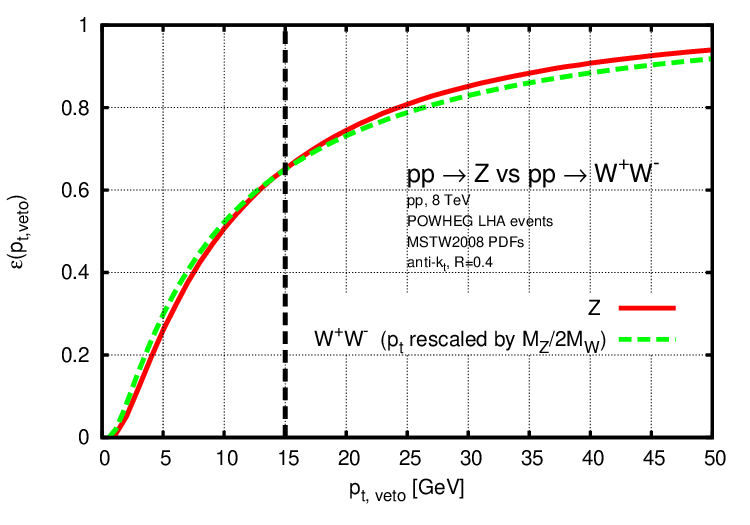}
  \hspace{-1em}
  \includegraphics[width=0.33\columnwidth]{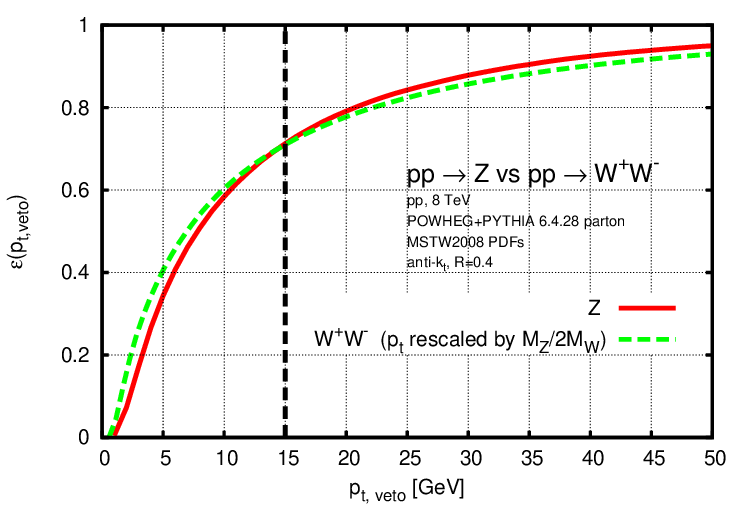}
  \hspace{-1em}
  \includegraphics[width=0.33\columnwidth]{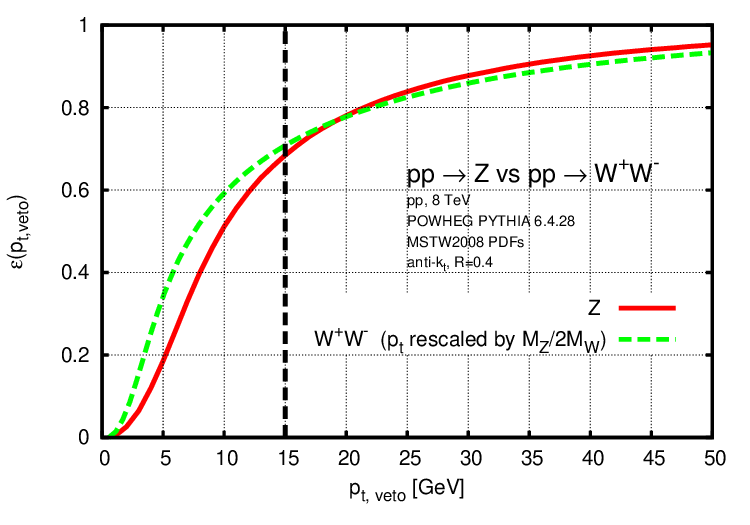}
  \caption{Jet-veto efficiency for $Z$ production obtained with the
    {\tt POWHEG BOX} and showered with {\tt Pythia} (v. 6.4.28 Perugia
    tune 350).}
  \label{fig:eff-PWG-Z}
\end{figure}

\begin{figure}[htp]
  \centering
  \includegraphics[width=0.5\columnwidth]{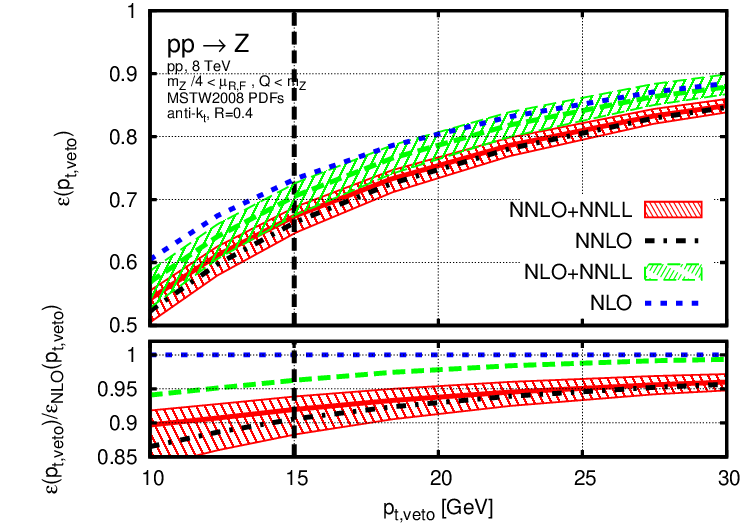}
    \hspace{-1em}
  \includegraphics[width=0.5\columnwidth]{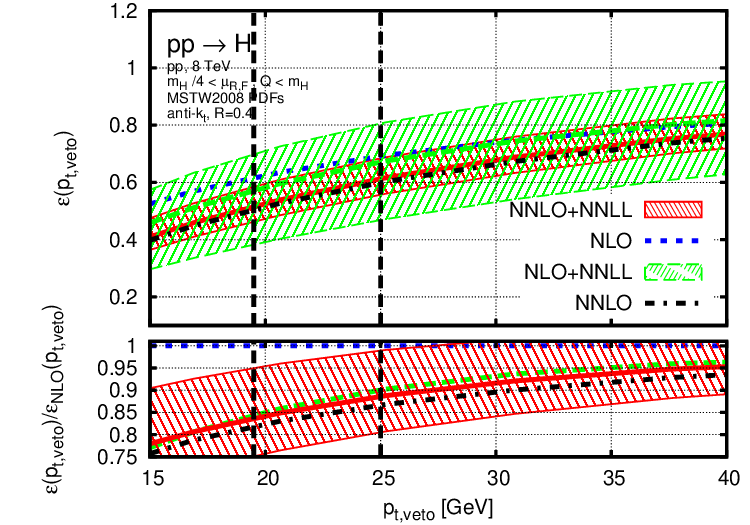}
  \caption{Jet-veto efficiency for single $Z$ (left) and $H$ (right)
    production. The uncertainty band for the resumed predictions is
    obtained by varying $\mu_R$, $\mu_F$, and the resummation scale
    $Q$ by a factor of two in either direction while keeping
    $1/2<\mu_R/\mu_F<2$. Moreover, the matching scheme for the
    jet-veto efficiency is also varied as shown in
    ref.~\cite{Banfi:2012jm}. The dashed black lines denote the
    $p_{\rm t,veto}$ values relevant for the present analysis (see
    text for more details).}
  \label{fig:eff-res}
\end{figure}

To study the impact of the jet veto, we consider the inclusive
jet-veto efficiency where no further cuts on leptons are applied. The
results are reported in Figure~\ref{fig:eff-PWG} for three different
{\tt Pythia} tunes. We observe that the suppression of the {\tt
  POHWEG} prediction with respect to the pure NLO at $p_{\rm t,veto}=
25$ GeV depends on the tune, and it varies in the range 6-9\%.
In order to check whether {\tt POWHEG} overestimates Sudakov
suppression effects, one can compare the Monte Carlo prediction to the
analytic resummation for the jet-veto efficiency.
Recently a NNLL resummation for the jet veto efficiency was carried
out for this process~\cite{Jaiswal:2014yba}, where it is shown that
the NNLL+NLO jet-veto efficiency is larger than the NLO one by a few
percent, however a matching of the NNLL and NNLO calculation has not
been done yet.
It is possible to provide an estimate of the NNLL+NNLO corrections by
looking at the jet-veto resummation for
Drell-Yan~\cite{Banfi:2012jm}. The difference between $Z$ and $W^+W^-$
production is mainly due to a different invariant mass for the
colourless final state, different parton luminosities, and the absence
of a $t$ channel in $Z$ production at the Born level.  As far as the
jet-veto {\it efficiency} is concerned, at small $p_{\rm t,veto}$,
where the logarithmic terms dominate, the main difference between the
two processes is due to the masses of the colourless final state
(which is the natural mass scale appearing in the large
logarithms). This is due to the fact that in the efficiency
differences in the virtual corrections and parton luminosities largely
cancel, while the real radiation pattern is the same in both
processes.
Moreover, the dependence on electroweak parameters also cancels in the
efficiency at small $p_{\rm t,veto}$. To convert the jet veto
efficiency from $Z$ to $W^+W^-$ production, neglecting for the moment
the $gg$ contribution, we impose that the argument of the large
logarithms is the same for both processes.  In both cases, the cross
section is integrated over the invariant mass of the colourless
final-state objects. Since the invariant mass spectrum is peaked at
$M_Z$ and $2 M_W$, respectively, and it steeply decreases for larger
invariant masses, we assume that the integral is dominated by the
value of the distribution at the peak and we thus consider the
respective masses as argument of the Sudakov logarithms.  This leads
to
\begin{equation}
\label{eq:rescaling}
\frac{ p_{\rm t,veto}^{DY}}{M_Z} = \frac{p_{\rm t,veto}^{WW}}{2 M_{W}}\,.
 \end{equation}
 Since ATLAS uses a veto of $p_{\rm t,veto}^{WW}=25$ GeV, one obtains
 $p_{\rm t,veto}^{DY}\sim$ 15 GeV. We emphasize that, because of this
 small $p_{\rm t,veto}$, the logarithmic terms are expected to
 dominate over the finite remainder. This correspondence can be tested
 with {\tt POWHEG} by comparing the jet-veto efficiency for the two
 processes. In order to study the Sudakov region in a
 shower-independent way, we perform the comparison at the Les Houches
 Event (LHE) level in Figure~\ref{fig:eff-PWG-Z}, where $p_{\rm
   t,veto}^{WW}$ has been rescaled by $M_{Z}/(2M_{W})$ according to
 Eq.~\ref{eq:rescaling}.
We see that after the rescaling the two efficiencies are in good
agreement in the small transverse momentum region.  
When the {\tt Pythia} parton shower effects (central plot in
Figure~\ref{fig:eff-PWG-Z}) are fully included the agreement remains,
although is slightly worse because of non-logarithmic corrections in
the parton shower. When also non-perturbative effects (including
hadronisation, multiple interactions, and intrinsic $p_t$ simulation)
are included, the above agreement is partly lost (right plot in
Figure~\ref{fig:eff-PWG-Z}), since some non-perturbative corrections
have a different scaling in $p_t$.
Still, we can use the relation between $p_{\rm t,veto}^{WW}$ and
$p_{\rm t,veto}^{DY}$ to estimate the impact of higher-order
logarithmic corrections on the jet-veto efficiency by looking at the
corresponding quantity in $Z$-boson production at $p_{\rm t,veto}= 15$
GeV, shown in Figure~\ref{fig:eff-res} (left).

By comparing the pure NLO (blue dashed line) and the NLO+NNLL (green
dashed line) calculation of Drell Yan at this transverse momentum
value, one observes a suppression by about 3-4\% when NNLL effects are
included. While the magnitude of the impact of the NNLL resummation is
found to be similar to that of ref.~\cite{Jaiswal:2014yba}, we observe
a reduction in the jet-veto efficiency rather than an
enhancement. This might be due to a different matching scheme and
treatment of higher-order corrections. 
In order to validate the estimate of Sudakov effects in {\tt POWHEG},
we can compare the lower panel in Figure~\ref{fig:eff-PWG} to the dashed green line in the lower panel of Figure~\ref{fig:eff-res},
representing the matching of the analytic resummation to NLO. In the case of
POWHEG, at $p_t = 25$ GeV, one finds a reduction in the efficiency that ranges between -6
and -9 \%. On the other hand NLO+NNLL results are just about -4\%
below the NLO result.
We thus observe that {\tt POWHEG} enhances considerably Sudakov effects with respect to the
NNLL+NLO result. On the
other hand, it is also clear from Fig.~\ref{fig:eff-res} that the
difference with respect to NLO is about -7-8\% when the NNLL is
matched to the NNLO result (red solid line). Hence, as far as jet-veto
effects are concerned, {\tt POWHEG} is accidentally close to the
NNLL+NNLO prediction at this veto scale. Therefore, the inclusive
cross section extrapolated using {\tt POWHEG} will be reasonably in
better agreement with the NNLO prediction, rather than with the NLO
one.

\begin{figure}[htp]
  \centering
  \includegraphics[width=0.5\columnwidth]{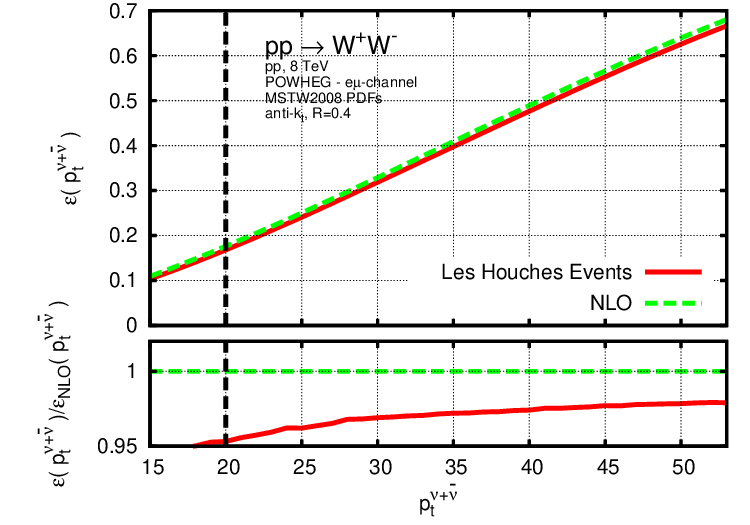}
  \hspace{-1em}
  \includegraphics[width=0.5\columnwidth]{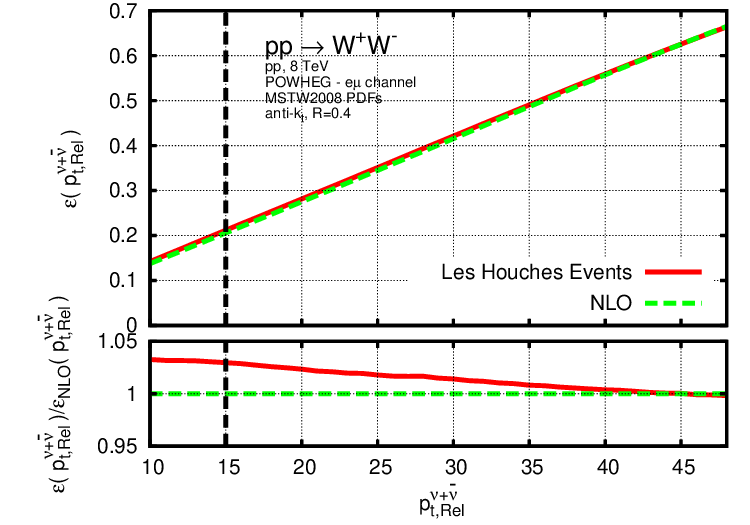}
  \caption{Efficiencies with a cut in either leading lepton's $p^{\nu+\bar\nu}_{t}$ (left) or $p^{\nu+\bar\nu}_{t,{\rm Rel}}$ (right). The plots report the comparison between LHE events and the pure NLO prediction, showing a systematic reduction at the LHA level for the actual fiducial cuts (black dashed lines).}
  \label{fig:lept-eff}
\end{figure}

A further reduction in the fiducial cross section is due to the way
the hardest emission is treated in {\tt POWHEG}, which was found to
slightly change the transverse momentum spectrum of the produced
leptons. As example, we show in Fig.~\ref{fig:lept-eff} the comparison
between the Les Houches events and the pure NLO for the missing
transverse momentum $p_{\rm t,miss}$ and the $p^{\nu+\bar\nu}_{t,{\rm
    Rel}}$ efficiencies.  We find that for the specific fiducial cuts
the overall effect amounts to the remaining $\sim$3\% reduction. The
latter effect is not enhanced further by the parton shower, which does
not change substantially the kinematics of leptons.
This difference is due to higher-order effects, however it leads to a
lower fiducial cross sections at LHE or parton-shower level in
comparison to pure NLO. This difference should be interpreted as a
systematic uncertainty associated with the Monte Carlo
generator.\footnote{It is possible to quantify this systematic
  uncertainty by varying in the {\tt POWHEG BOX} code the parameter
  {\tt hdamp}~\cite{Nason:2012pr}, which controls the amount of real
  radiation included in the Sudakov form factor. In this note, we do not study 
  this dependence and perform the whole analysis using the default settings, 
  i.e. without setting the {\tt hdamp} parameter.}

We can provide an estimate for the fiducial cross section using the
information on the NNLL+NNLO jet veto efficiency, and on the NNLO
inclusive cross section, and assuming that QCD higher order
corrections do not affect the lepton acceptances.  The resulting
fiducial cross section can be expressed as
\begin{equation}
\label{eq:fid-est}
\sigma_{\rm fid.}^{\rm th.} = \sum_{c \,\in \,{\rm channel}}\sigma_{\rm fid.}^{(c),{\rm NLO}}\times\frac{\epsilon^{(c),{\rm NNLL+NNLO}}(p^{(c)}_{\rm t,veto})}{\epsilon^{(c),{\rm NLO}}(p^{(c)}_{\rm t,veto})}\times\frac{\sigma_{\rm incl.}^{(c),{\rm NNLO}}}{\sigma_{\rm incl.}^{(c),{\rm NLO}}},
\end{equation}
where the sum is performed over the three different channels,
i.e. $q\bar q\to W^+W^-$, $gg\to W^+W^-$, and $gg\to H\to W^+W^-$.
Using Figs.~\ref{fig:eff-res}, one can give an estimate of the jet
veto efficiencies at the three relevant $p_{\rm t,veto}^{(c)}$
scales. For the $q\bar q\to W^+W^-$ channel, we consider the Drell-Yan
efficiencies at $p_{\rm t,veto}^{q\bar q}=15$\,GeV, while for the
$gg\to W^+W^-$ and $gg\to H\to W^+W^-$ channels we consider the Higgs
production efficiencies at $p_{\rm t,veto}^{gg\to WW}=19.5$\,GeV, and
$p_{\rm t,veto}^{gg\to H\to WW}=25$\,GeV, respectively.  The NLO cross
sections are taken from {\tt MCFM}, while the NNLO cross sections are
taken from ref.~\cite{Gehrmann:2014fva}.
We obtain the fiducial cross sections reported in
Table~\ref{tab:inclusive}.  The theory uncertainties have been
obtained by combining in quadrature the symmetrized errors in the {\tt
  MCFM} cross sections, estimated jet-veto efficiencies, and NNLO
total cross section, respectively. For a more reliable estimate of the
uncertainty, one could use one of the methods available in the
literature~\cite{Stewart:2011cf,Banfi:2012jm,Boughezal:2013oha}.  We
observe that the estimated fiducial cross sections are in very good
agreement with data. Therefore, also the extrapolated inclusive cross
section will be in agreement with the NNLO prediction.

\begin{table}[htp]
\centering
\footnotesize
\begin{tabular}{|c||c|c|}
\hline
   decay mode  & $\sigma_{\rm fid.}^{\rm exp.}$\,[fb] & $\sigma_{\rm fid.}^{\rm th.}$\,[fb] \\
 \hline
$\emu$ &$377.8^{+6.9}_{-6.8}{\rm (stat.)}^{+25.1}_{-22.2}{\rm (syst.)}^{+11.4}_{-10.7}{\rm (lumi.)} $& $353.5^{+15.5}_{-15.5}$\\
$\ee$ &$68.5^{+4.2}_{-4.1}{\rm (stat.)}^{+7.7}_{-6.6}{\rm
  (syst.)}^{+2.1}_{-2.0}{\rm (lumi.)} $& $68.1^{+2.9}_{-2.9}$\\
$\mumu$  &$74.4^{+3.3}_{-3.2}{\rm (stat.)}^{+7.0}_{-6.0}{\rm
  (syst.)}^{+2.3}_{-2.1}{\rm (lumi.)} $ & $74.1^{+3.2}_{-3.2}$\\
\hline 
\end{tabular}
\caption{Comparison between the measured fiducial cross section and the theory prediction with estimated NNLL+NNLO effects. Theory uncertainties have been symmetrized and combined in quadrature.
  \label{tab:inclusive}}
\end{table}

\section{Conclusions}
Despite the substantial tension with the NLO prediction reported by
ATLAS for the inclusive $W^+W^-$ cross section at 8 TeV, our findings
show that the measured fiducial cross sections are in good agreement
(at about the 1-$\sigma$ level) with the NLO result for all leptonic
decay modes.

The discrepancy observed by ATLAS at 8 TeV can be explained to a large
extent by studying the Monte Carlo predictions used in the
extrapolation to the inclusive cross section.
The prediction obtained with the {\tt POWHEG BOX} reduces the NLO
result by 9-11\%.  The Monte Carlo result for the jet-veto efficiency
is found to overestimate Sudakov suppression effects with respect to
the analytic resummation.
Furthermore, at the Les Houches level, the leptons' transverse
momentum spectrum is slightly different than the NLO distribution.
This is a higher-order effect, and it leads to a systematic reduction
in the fiducial cross section (of about 3\% in the present case) at
the parton-shower level with respect to NLO. As a result, the fiducial
cross-section is reduced and the extrapolated ("measured'') inclusive
cross section tends to be overestimated.  The latter effect should be
taken into account as a potential additional systematic uncertainty
associated with the matching procedure of higher-order calculations to
a parton shower.

Using the NNLL+NNLO predictions for Drell-Yan and Higgs production, we
provide an estimate for the fiducial cross section, including these
higher-order effects. It suggests that the agreement found at NLO for
the fiducial cross sections will remain good, or even improve once
NNLL+NNLO effects are included in the theory predictions.
A better assessment of higher order effects and related uncertainties
is today possible by matching the recently computed NNLO
corrections~\cite{Gehrmann:2014fva} to the existing NNLL
resummation~\cite{Jaiswal:2014yba}.
It would be also useful to encode the NNLO prediction in a fully
exclusive NNLO + parton shower generator, on the line of what has been
recently done for Higgs-~\cite{Hamilton:2013fea,Hoche:2014dla} and
Drell-Yan
production~\cite{Karlberg:2014qua,Hoeche:2014aia}.\footnote{In
  ref.~\cite{Hoche:2014dla} the authors report problems in describing
  the low $p_t$ region with their NNLO parton shower approach.}
Further studies are also required for the formally NNLO $gg$-initiated
channel. At lowest order it amounts to 3-4\% of the NLO cross section
at the inclusive level (i.e. of the same order of the Higgs-mediated
reaction), but it raises to 8-9\% after fiducial cuts are taken into
account since the jet veto is ineffective when applied to the tree
level prediction. Due to the large amount of initial state radiation,
this channel will experience a sizable Sudakov suppression, and has a
potentially large K factor.  We estimate the NNLL+NNLO effects in the
jet-veto efficiency, but a more detailed study is necessary.

\section*{Acknowledgements}
We would like to thank and P.~Nason and G.~Salam for valuable
discussions, and A.~Banfi and K.~Hamilton for useful comments on the
manuscript.
GZ is supported by the ERC grant 614577. PM is supported by the
Research Executive Agency (REA) of the European Union under the Grant
Agreement number PITN-GA-2010-264564 (LHCPhenoNet), and by the Swiss
National Science Foundation (SNF) under the grant PBZHP2-147297.
This research was partly supported by the Munich Institute for Astro-
and Particle Physics (MIAPP) of the DFG cluster of excellence "Origin
and Structure of the Universe" (PM and GZ) and by the Mainz Institute
for Theoretical Physics (MITP) of the PRISMA excellence cluster (GZ).
We would like to thank the Galileo Galilei Institute (PM and GZ) and
CERN (PM) for hospitality while part of this work was carried
out.

\section*{Note added}
\begin{figure}[htp]
  \centering
  \includegraphics[width=0.5\columnwidth]{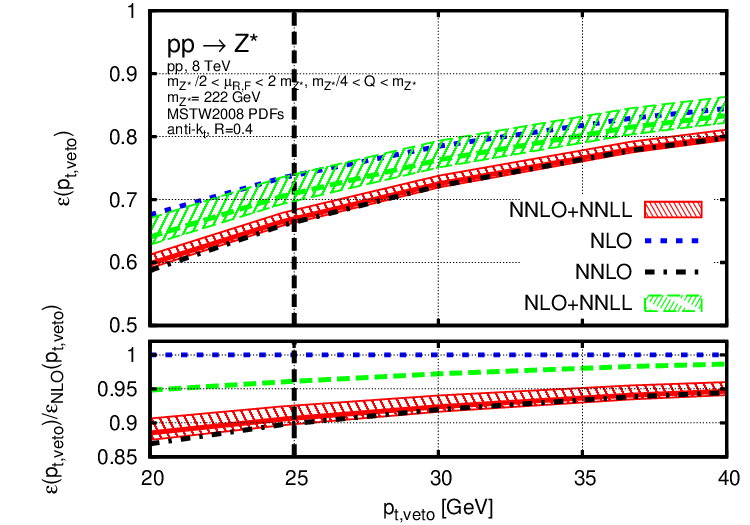}
    \hspace{-1em}
  \includegraphics[width=0.5\columnwidth]{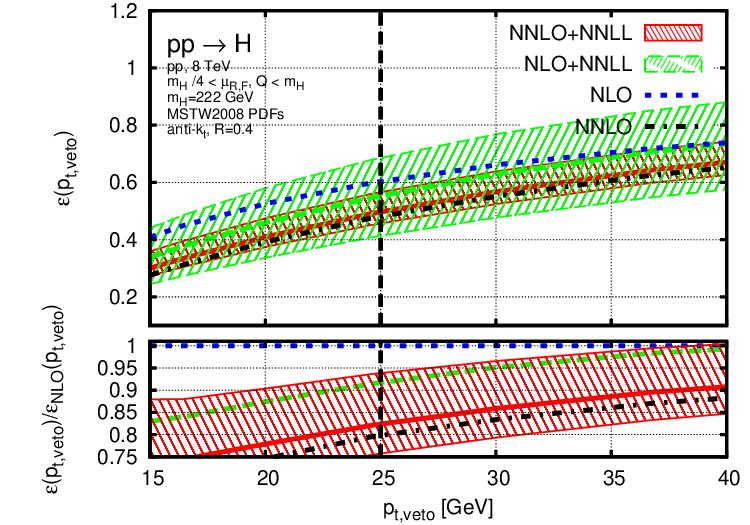}
  \caption{As in Figure~\ref{fig:eff-res}, but the jet-veto efficiency
    for single $Z$ (left) and $H$ (right) production are both obtained
    with invariant mass $M=222$\, GeV. The dashed black line at
    $p_{\rm t,veto}=25$\,GeV denotes the value relevant for the $q\bar
    q \to W^+W^-$ and $gg\to W^+W^-$ in
    Eq.~\eqref{eq:fid-est}. For the $gg\to H\to W^+W^-$ the
    result is unchanged.}
  \label{fig:eff-res-BFNR}
\end{figure}

After this article appeared, the NNLL+NLO jet-veto cross section for
generic EW boson production was studied in
reference~\cite{Becher:2014aya}. The authors pointed out that the
rescaling~\eqref{eq:rescaling} is only approximate. We are in fact
aware of this. In particular, in addition to the
rescaling~\eqref{eq:rescaling}, one should also replace the
renormalization scale at which the strong coupling is evaluated with
the new value for $W^+W^-$ production, and change the values of
$\mu_F$ and $M^2$ (i.e. the squared invariant mass of the colour
singlet system) in the parton luminosities.  Our approximation relies
on the fact that the effect of the scale change in the strong coupling is
moderate, while the effects of parton luminosities largely cancel in
the ratio of efficiencies in our formula~\eqref{eq:fid-est}. 
Reference~\cite{Becher:2014aya} suggests to change the DY invariant
mass in $Z$-boson production from $M_Z$ to $222$\,GeV (i.e. the median
of the $W^+W^-$ invariant mass distribution) in order to relate the
$Z$ production efficiency to the $W^+W^-$ one. They have verified that
with this prescription the DY efficiency at NNLL is in excellent
agreement with the actual $W^+W^-$ result. We have done this exercise,
and the resulting plots analogous to our Figure~\ref{fig:eff-res} are
shown in Figure~\ref{fig:eff-res-BFNR}.

We observe that, as expected, the ratios of efficiencies to their NLO
value are almost unchanged. We have used these new values to reproduce
the approximate fiducial cross sections of
Table~\ref{tab:inclusive}. The new results are reported in
Table~\ref{tab:inclusive-BFNR}, and we do not observe any significant
difference from our estimate.

\begin{table}[htp]
\centering
\footnotesize
\begin{tabular}{|c||c|c|}
\hline
   decay mode  & $\sigma_{\rm fid.}^{\rm exp.}$\,[fb] & $\sigma_{\rm fid.}^{\rm th.}$\,[fb] \\
 \hline
$\emu$ &$377.8^{+6.9}_{-6.8}{\rm (stat.)}^{+25.1}_{-22.2}{\rm (syst.)}^{+11.4}_{-10.7}{\rm (lumi.)} $& $349.8^{+11.8}_{-11.8}$\\
$\ee$ &$68.5^{+4.2}_{-4.1}{\rm (stat.)}^{+7.7}_{-6.6}{\rm
  (syst.)}^{+2.1}_{-2.0}{\rm (lumi.)} $& $67.4^{+2.2}_{-2.2}$\\
$\mumu$  &$74.4^{+3.3}_{-3.2}{\rm (stat.)}^{+7.0}_{-6.0}{\rm
  (syst.)}^{+2.3}_{-2.1}{\rm (lumi.)} $ & $73.3^{+2.4}_{-2.4}$\\
\hline 
\end{tabular}
\caption{As in Table~\ref{tab:inclusive}, but using efficiencies
  ratios from Figure~\ref{fig:eff-res-BFNR} for $q\bar q\to W^+W^-$
  and $gg\to W^+W^-$ contributions.
  \label{tab:inclusive-BFNR}}
\end{table}

\end{document}